\documentclass{ws-acs}
\usepackage{url}

\begin{document}

\makeatletter
\def\@biblabel#1{[#1]}
\makeatother

\markboth{T.Conlon, H.J.Ruskin, M.Crane}{Multiscaled Cross-Correlation Dynamics in Financial Time-Series}

\title{Multiscaled Cross-Correlation Dynamics in Financial Time-Series}

\author{T. CONLON\footnotemark, H.J. RUSKIN and M. CRANE}

\address{School of Computing, Dublin City University, Ireland\footnote{\footnotesize School of Computing, Dublin City University, Glasnevin, Dublin 9, Ireland}\\
*tconlon@computing.dcu.ie}

\maketitle

\begin{history}
\received{06 October 2008}
\end{history}

\begin{abstract}
The cross-correlation matrix between equities comprises multiple interactions between traders with varying strategies and time horizons.  In this paper, we use the Maximum Overlap Discrete Wavelet Transform (MODWT) to calculate correlation matrices over different time scales and then explore the \textit{eigenvalue spectrum} over sliding time windows. The dynamics of the eigenvalue spectrum at different times and scales provides insight into the interactions between the numerous constituents involved. 

Eigenvalue dynamics are examined for both medium and high-frequency equity returns, with the associated correlation structure shown to be dependent on both time and scale.  Additionally, the \emph{Epps} effect is established using this multivariate method and analysed at longer scales than previously studied.  A partition of the eigenvalue time-series demonstrates, at very short scales, the emergence of negative returns when the largest eigenvalue is greatest.  Finally, a portfolio optimisation shows the importance of time-scale information in the context of risk management.

\end{abstract}

\section{Introduction}
\label{Intro}
In recent years, the equal-time cross-correlation matrix has been studied extensively for a variety of multivariate data sets across different disciplines such as financial data \cite{Laloux_1999,Plerou_1999,Laloux_2000,Plerou_2000,Gopikrishnan_2000,Utsuki_2004,Wilcox_2004,Sharifi_2004,Conlon_2007,Conlon_2008,Conlon_2009}, electroencephalographic (EEG) recordings \cite{Muller_2005,Muller_2006_b,Schindler_2007}, magnetoencephalographic (MEG) recordings \cite{Kwapien_2000} and others.  In particular, Random Matrix Theory (RMT) has been applied to filter the relevant information from the statistical fluctuations inherent in empirical cross-correlation matrices, constructed for various types of financial, \cite{Laloux_1999,Plerou_1999,Laloux_2000,Plerou_2000,Gopikrishnan_2000,Utsuki_2004,Wilcox_2004,Sharifi_2004,Conlon_2007}, and MEG, \cite{Kwapien_2000}, time-series.  In this paper, we extend previous Complex Systems research, by examining time and scale dependent dynamics of the correlation matrix, in order to better understand the ever-changing level of synchronisation between financial time-series.

The dynamics of the largest eigenvalue of a cross-correlation matrix over small windows of time was studied, \cite{Drozdz_2000}, for the Dow Jones Industrial Average (DJIA) and DAX  indices.  This analysis revealed evidence of time-dependence between `drawdowns'/`drawups' and an increase/decrease in the largest eigenvalue. The interactions between stocks of two different markets (DAX and DJIA) were then investigated, \cite{Drozdz_2001}, revealing two distinct eigenvalues of the combined cross-correlation matrix, corresponding to each market.  Adjusting for time-zone delays, the two eigenvalues coincided, implying that one market leads the dynamics in the other.   

It has been suggested recently by several authors, \cite{Guhr_2003,Burda_2004_a,Burda_2004b,Malevergne,Kwapien_2006} that there may, in fact, be some real correlation information hidden in the RMT defined part of the eigenvalue spectrum.  A multivariate technique involving the equal-time cross-correlation matrix has been shown, \cite{Muller_2005}, to characterise dynamical changes in nonstationary multivariate time-series.  Using this technique, the authors also demonstrated how correlation information can be observed in the noise band.  As the synchronisation of $k$ time-series within an $M-$dimensional multivariate time-series increases, a repulsion between eigenstates of the correlation matrix results, in which $k$ levels participate.  Through the use of artificially-created time-series with pre-defined correlation dynamics, it was demonstrated that there exist situations, where the relative change in eigenvalues from the lower edge of the spectrum is greater than that for the large eigenvalues, implying that information drawn from the smaller eigenvalues is highly relevant.  

This technique was subsequently applied to the dynamic analysis of the eigenvalue spectrum of the equal-time cross-correlation matrix of multivariate Epileptic Seizure time-series, using sliding windows.  The authors demonstrated that information about the correlation dynamics is visible in \emph{both} the lower and upper eigenstates.  A further study, \cite{Schindler_2007}, which investigated temporal dynamics of focal onset epileptic seizures\footnote{A focal onset or partial seizure occurs when the discharge starts in one area of the brain and then spreads over other areas.}, showed that the zero-lag correlations between multichannel EEG signals tend to decrease during the first half of a seizure and increase gradually before the seizure ends.  Information about cross correlations was also found in the RMT bulk of eigenvalues, \cite{Muller_2006_b}, with that extracted at the \emph{lower} edge statistically \emph{more significant} than that from the larger eigenvalues.  Application of this technique to multichannel EEG data showed small eigenvalues to be more sensitive to detection of subtle changes in the brain dynamics than the largest.  A time and frequency based approach, has recently demonstrated the importance of correlation dynamics at high-frequencies during seizure activity, \cite{Conlon_CBM}.

Although originally applied to complex EEG seizure time-series, \cite{Muller_2005,Muller_2006_b,Schindler_2007}, these techniques were adapted to financial time-series, \cite{Conlon_2009}, where the correlation dynamics for medium-frequency data were calculated using short time-windows.  Changes in the correlation structure were shown to be visible in both large and small eigenvalue dynamics.  Further, `drawdowns' and `drawups' were shown to be market dependent, with corresponding characteristic changes in relative eigenvalue size found at both ends of the eigenvalue spectrum.  This analysis was carried out using daily data, with the effects of varying data granularity not examined.  In this paper, we build upon this analysis by studying the multiscale effects on correlation dynamics using medium and high-frequency data.

The correlation structure for various markets was studied in detail, \cite{Marsili_2007}, for data encompassing a number of time horizons ranging from $5$ to $255$ minutes.  By removing the centre of mass, (the market), the authors found that the correlation structure is well defined at high frequency, while for the original data the structure emerges at longer time horizons, (using stocks from the New York Stock Exchange).


Wavelet multiscale analysis has been used by numerous authors to decompose economic and financial time-series into orthogonal time-scale components of varying granularities.  Gen\c{c}ay and co-workers have variously examined the scaling properties of foreign exchange volatility, \cite{Gencay_2001b}, and volatility models without intraday seasonalities, \cite{Gencay_2001c}, using wavelet multiscaling techniques.  The systematic risk in a Capital Asset Pricing Model was estimated over different granularities, \cite{Gencay_2003}, where it was shown that the return of a portfolio and its Beta\footnote{The Beta of an asset is a measure of the volatility, or systematic risk, of a security or a portfolio in comparison to the market as a whole.} become stronger as the scale increases for the S\&P 500. 

An extensively studied characteristic of stock market behaviour, is the increase of stock return cross-correlations as the sampling time scale increases, a phenomenon known as the \textit{Epps} effect, \cite{Epps}.  More recently, analysis of time-dependent correlations between high-frequency stocks demonstrated, however, that market reaction times have increased due to greater efficiency, \cite{Kertesz_2006}.  A diminution of the Epps effect with time is one consequence of increased market efficiency.  Trading asynchronicity was demonstrated to be not solely responsible for the effect, \cite{Kertesz_2007_a}, with the characteristic time apparently independent of the trading frequency. Further analysis using a toy model of Brownian motion and memoryless renewal process, \cite{Kertesz_2007_b}, found an exact expression for the Epps frequency dependence, with reasonable fitting also for empirical data.  In fact, the effect was shown, \cite{Kertesz_2007_c}, not to scale with market activity but to be due to reaction times, rather than market activity.

In this paper, we extend the multivariate correlation technique, first applied to complex EEG seizure data, \cite{Muller_2005,Muller_2006_b,Schindler_2007}, and later to medium frequency financial data, \cite{Conlon_2009}.  This interdisciplinary approach, allows characterisation of correlation changes in time.  By decomposing equity returns into their component scales in short time-windows, we are able to study the correlation and associated eigenspectrum at various scales, allowing a time-scale analysis of the eigenspectrum.  Using both medium and high-frequency stock returns, we examine the dynamics of the large eigenvalues and demonstrate the time-scale dependence of the correlation structure.  We show how the Epps effect, \cite{Epps}, can be demonstrated through a multivariate approach and expand previous work by examining considerably longer scales.  Finally, we look at some applications of the work, with scale dependent risk characterisation and portfolio optimisation examined.  Sections (\ref{methods} \verb`-` \ref{data}) describe the techniques and data used, Section (\ref{results}) details the results obtained and conclusions are given in Section (\ref{conclusions}).

\section{Methods}
\label{methods}
\subsection{Wavelet multiscale analysis}
\label{MODWT}
Wavelets provide an efficient means of studying the multiresolution properties of a signal, allowing decomposition into different time horizons or frequency components (Discrete Wavelet Transform, DWT), \cite{Percival_2000,Gencay_2001}.  The definitions of the two basic wavelet functions, the father $\phi$ and mother $\psi$ wavelets are given as:
\begin{eqnarray}
\phi_{j,k}\left(t\right) = 2^{\frac{j}{2}} \phi\left(2^{j}t - k\right) \\
\psi_{j,k}\left(t\right) = 2^{\frac{j}{2}} \psi\left(2^{j}t - k\right) 
\end{eqnarray}
where $j = 1, \ldots J$ in a $J$-level decomposition.  The father wavelet integrates to 1 and reconstructs the longest time-scale component of the series, while the mother wavelet integrates to 0 and is used to describe the deviations from the trend.  The wavelet representation of a discrete signal $f(t)$ in $L^{2}(R)$ is:
\begin{eqnarray}
f(t) & = & \sum_{k}s_{J,k}\phi_{J,k}(t) + \sum_{k}d_{J,k}\phi_{J,k}(t)+ \ldots  + \sum_{k}d_{1,k}\phi_{1,k}(t)
\end{eqnarray}
where $J$ is the number of multiresolution levels (or \emph{scales}) and $k$ ranges from $1$ to the number of coefficients in the specified level. The coefficients $s_{J,k}$ and $d_{J,k}$ are the smooth and detail component coefficients respectively and given by
\begin{eqnarray}
s_{J,k} = \int \phi_{J,k} f(t)dt &\   	& \\
d_{j,k} = \int \psi_{j,k} f(t)dt &\  \	  &   	(j = 1, \ldots J)
\end{eqnarray}
Each of the coefficient sets $S_{J},d_J,d_{J-1}, \ldots d_1$ is called a \emph{crystal}.

In this paper, we apply the Maximum Overlap Discrete Wavelet Transform (MODWT), \cite{Percival_2000,Gencay_2001}, a linear filter that transforms a series into coefficients related to variations over a set of scales.  Like the DWT it produces a set of time-dependent wavelet and scaling coefficients with basis vectors associated with a location $t$ and a unitless scale $\tau_{j} = 2^{j-1}$ for each decomposition level $j = 1, \ldots, J_{0}$ .  The MODWT, unlike the DWT, has a high level of redundancy, however, is nonorthogonal and can handle any sample size $N$.  It retains downsampled\footnote{Downsampling or decimation of the wavelet coefficients retains half of the number of coefficients that were retained at the previous scale and is applied in the DWT} values at each level of the decomposition that would be discarded by the DWT. This reduces the tendency for larger errors at lower frequencies when calculating frequency dependent variance and correlations, (Section \ref{Wavelet_Var}), as more data is available.  Having decomposed the equity returns into their component time-scales, we then calculate the correlation between these components.

\subsection{Correlation dynamics}
The equal-time correlation matrix between time-series of stock returns is calculated using a sliding time window where the number of stocks, $N$, is smaller than the window size $T$.  Given time-series of stock returns $R_{i} \left(t\right)$, $i = 1, \ldots,N$, we normalise the time-series within each window as follows:
\begin{equation}
r_{i}\left(t\right) = \frac{R_{i} \left(t\right) - \widehat{R_{i} \left(t\right)}}{\sigma_{i}}
\end{equation}
where $\sigma_{i}$ is the standard deviation of stock $i = 1, \ldots,N$ and $\widehat{R_{i}}$ is the time average of $R_{i}$ over a time window of size $T$. Then, the equal time cross-correlation matrix, expressed in terms of $r_{i}\left(t\right)$, is
\begin{equation}
C_{ij} \equiv \left\langle r_{i}\left(t\right) r_{j}\left(t\right) \right\rangle
\label{cross_corr}
\end{equation}
The elements of $C_{ij}$ are limited to the domain $-1 \leq C_{ij} \leq 1$, where $C_{ij} = 1$ defines perfect positive correlation, $C_{ij} = -1$ corresponds to perfect negative correlation and $C_{ij} = 0$ corresponds to no correlation. In matrix notation, the correlation is expressed as $\mathbf{C} = \frac{1}{T} \mathbf{RR}^{t}$, where $\bf{R}$ is an $N\times T$ matrix with elements $r_{it}$.

The eigenvalues $\mathbf{\lambda}_{i}$ and eigenvectors $\mathbf{\hat{v}}_{i}$ of the correlation matrix $\mathbf{C}$ are found using $\mathbf{C} \mathbf{\hat{v}}_{i} = \mathbf{\lambda}_{i} \mathbf{\hat{v}}_{i}$ and then ordered according to size, such that $\mathbf{\lambda}_{1} \leq \mathbf{\lambda}_{2}\leq \ldots \leq \mathbf{\lambda}_{N}$.  Given that the sum of the diagonal elements of a matrix (the Trace) remains constant under a linear transformation,  $\sum_{i} \lambda_i$ must always equal the trace of the original correlation matrix.  Hence, if some eigenvalues increase then others must decrease, to compensate, and vice versa (\textit{Eigenvalue Repulsion}).  

There are two limiting cases for the distribution of the eigenvalues \cite{Muller_2005,Schindler_2007}, with perfect correlation, $C_{i} \approx 1$, when the largest is maximised with value $N$ (all others taking value zero).  When each time-series consists of random numbers with average correlation $C_{i} \approx 0$, the corresponding eigenvalues are distributed around $1$, (where deviations are due to spurious random correlations).  Between these two extremes, the eigenvalues at the lower end of the spectrum can be much smaller than $\lambda_{max}$.  By calculating the eigenspectrum associated with the correlation matrix at different scales in each time window, we get time-series of eigenvalues $\lambda_{mn}({\tau})$, where $m$ denotes the eigenvalue number and $n$ is the scale studied in a particular time-window, $\tau$.

To study the market behaviour when eigenvalues take extreme values, we need to partition the later depending on their size.  This is achieved by first normalising each eigenvalue in time using
\begin{equation}
\mathbf{\tilde{\lambda}}_{i}(t) = \frac{\left(\mathbf{\lambda}_{i} - \mathbf{\bar{\lambda}(\tau)}\right)}{\sigma^{\lambda(\tau)}}
\label{normalise}
\end{equation}
where $\mathbf{\bar{\lambda}(\tau)}$ and $\sigma^{\lambda(\tau)}$ are the mean and standard deviation of the eigenvalues over a particular reference period, $\tau$.  By expressing the eigenvalue time-series in terms of standard deviation units (SDU), we can then easily partition each according to it's relative size.

Combining the techniques of multivariate correlation analysis and multiscaling, we are able to provide some novel insight into the dependence of correlation on both time and scale.  To enable this analysis, we first show how correlations are calculated using wavelet coefficients.

\subsection{Wavelet covariance and correlation}
\label{Wavelet_Var}
The wavelet covariance between functions $f(t)$ and $g(t)$ is defined to be the covariance of the wavelet coefficients at a given scale.  The \emph{unbiased} estimator of the wavelet covariance at the $j^{th}$ scale is given by
\begin{equation}
\nu_{fg}(\tau_j) = \frac{1}{M_j} \sum^{N-1}_{t = L_{j}-1} \tilde{D}_{j,t}^{f(t)} \tilde{D}_{j,t}^{g(t)}
\end{equation}
where all the wavelet coefficients affected by the boundary are removed \cite{Percival_2000} and $M_j = N - L_j +1$.  The wavelet variance at a particular scale $\nu^{2}_{f} (\tau_j)$ is found similarly. 

The MODWT estimate of the wavelet cross correlation between functions $f(t)$ and $g(t)$ may then be calculated using the wavelet covariance and the square root of the wavelet variance of the functions at each scale $j$.  The MODWT estimator \cite{Gencay_2001} of the wavelet correlation is then given by:
\begin{equation}
\label{Wave_Corr}
\rho_{fg} (\tau_j) = \frac{\nu_{fg}(\tau_j)}{\nu_{f}(\tau_j)\nu_{g}(\tau_j)}
\end{equation}

\section{Data}
\label{data}
\begin{figure}[htbp!]
\begin{center}
\includegraphics[height=78mm,width=135mm]{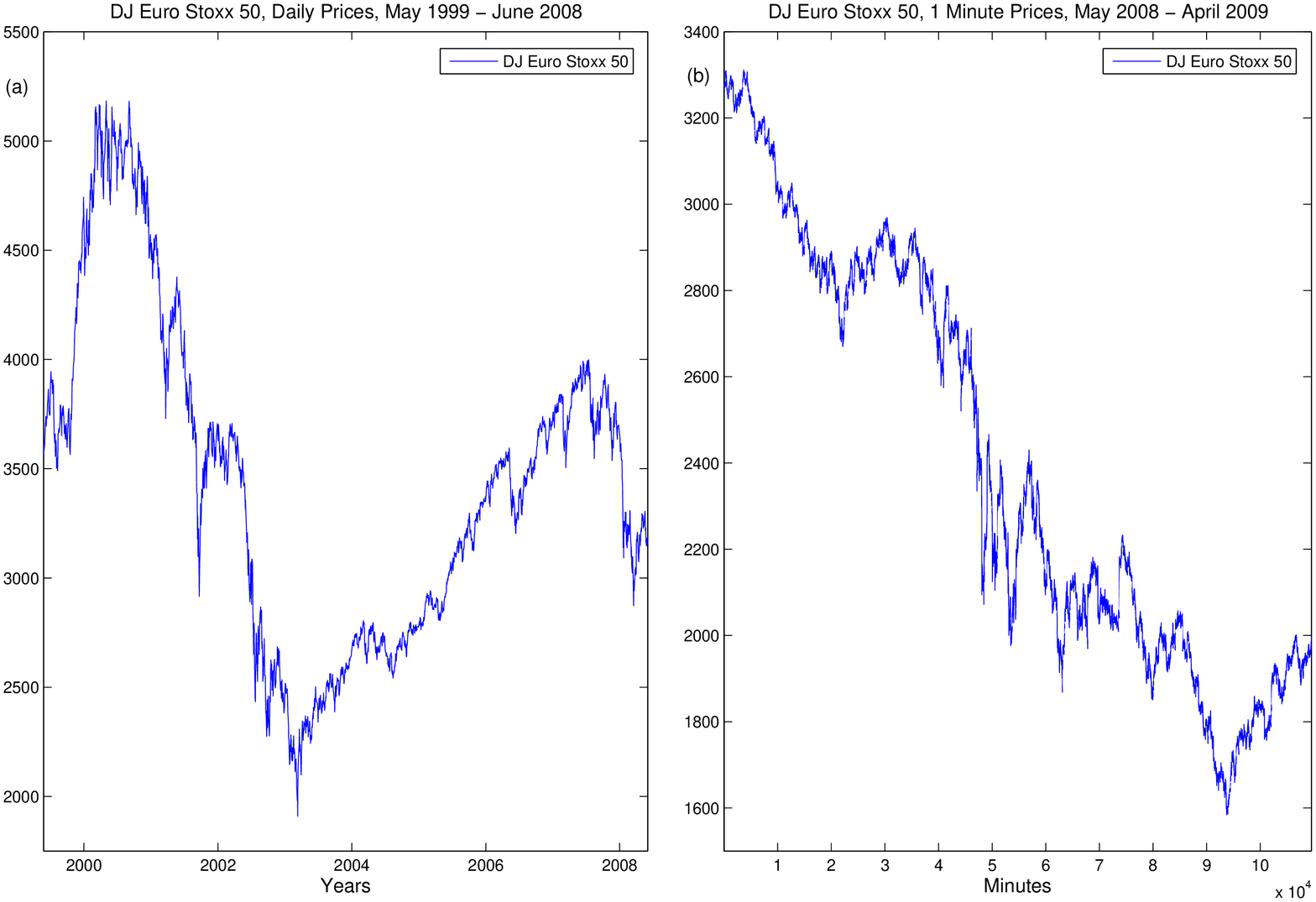}
\caption{a) DJ Euro Stoxx 50, Daily Prices May $1999$ - June $2008$ (b) DJ Euro Stoxx 50, $1$ Minute Prices May $2008$ - April $2009$}
\label{indexPrices}
\end{center}
\end{figure}

The initial data set comprises the 49 equities of the Dow Jones (DJ) Euro Stoxx 50 where full price data is available from May $1999$ to August $2007$, resulting in $2183$ daily returns, Fig.~\ref{indexPrices}(a).  The Dow Jones Euro Stoxx 50 is a stock index of Eurozone equities, designed to provide a blue-chip representation of Eurozone supersector leaders.  The small number of stocks in the index allows calculation of the cross-correlation matrices for small time windows, without reducing the matrix rank.

The second data set studied consists of high-frequency returns for the Dow Jones (DJ) Euro Stoxx 50 from May $2008$ to April $2009$, resulting in $109,540$ one-minute Returns, Fig.~\ref{indexPrices}(b).  High-frequency data allows us to examine additional features and analyse whether the correlation dynamics previously found, \cite{Conlon_2009}, are inherent at all time-scales or if there is a gradual emergence.  The time-frame studied is of particular interest, with sustained market drops and high levels of volatility.  Understanding market interactions in a period such as this may assist in forming portfolios that are robust to large downwards moves.

\section{Results}
\label{results}
To decompose the data into component time-scales, we selected the least asymmetric (LA) wavelet, (known as the Symmlet), which exhibits near symmetry about the filter midpoint.  LA filters are available in even widths and the optimal filter width is dependent on the characteristics of the signal and the length of the data.  The filter width chosen for this study was the LA8, (where the 8 refers to the width of the scaling function) since it enabled us to accurately calculate wavelet correlations to relatively long scales, dependent upon the time window used.  Although the MODWT can accommodate any level $J_0$, in practise the largest level is chosen so as to prevent decomposition at scales longer than the total length of the data series.

\subsection{Medium Frequency Eigenvalue dynamics}
\label{MedFreqDyn}
We first focus on medium-frequency equity returns measured at a one-day horizon.  For each stock, using a sliding time-window of $100$ days, we decompose the returns into their component scales in each window, using the wavelet transform.   The correlation between the wavelet coefficients, corresponding to each equity, is then calculated at each scale, as described (Section~\ref{Wavelet_Var}).  Time-series of eigenvalues are found at each scale, by calculating the associated eigenspectrum.  This approach of calculating the eigenspectrum in sliding time-windows, allows analysis of the dependence of correlation dynamics on granularity.

\begin{figure}[htbp!]
\begin{center}
\includegraphics[height=82mm,width=135mm]{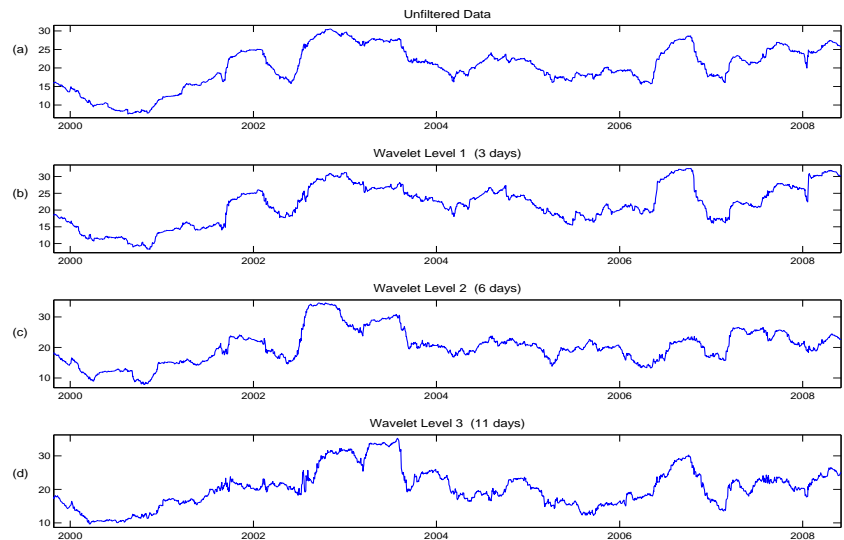}
\caption{a) Largest eigenvalue dynamics original data (b) 3 day scale (c) 6 day scale (d) 11 day scale}
\label{eValueLarge}
\end{center}
\end{figure}

The time-window was chosen such that $Q = \frac{T}{N} = 2.04$, thus ensuring that the data would be close to stationary in each sliding window, (different values of $Q$ were examined, \cite{Conlon_2009}, with little variation in the results). Fig.~\ref{indexPrices}(a) shows the value of the DJ Euro Stoxx Index over the period studied.  Fig.~\ref{eValueLarge}(a) displays the largest eigenvalue, calculated using the unfiltered (one day) time-series data for sliding time windows.  As shown, the largest eigenvalue is far from static, rising from a minimum of $7.5$ to a maximum of $30.5$ from early $2001$ to late $2003$, (coinciding with the bursting of the ``tech'' bubble).  This corresponds to an increase in the influence of the ``Market'', with the behaviour of traders becoming more correlated.  The next major increase occurred in early $2006$, followed by a relatively marked decline until the beginning of the ``Credit Crunch'' in $2007$.  Similar to~\cite{Drozdz_2000}, we note an increase in the value of the largest eigenvalue during times of market stress, with lower values during more ``normal'' periods.

We next calculate, using the MODWT (Section~\ref{MODWT}), the value of the largest eigenvalue of the cross-correlation matrix over longer time horizons of $3$, $6$ and $11$ days (Fig.~\ref{eValueLarge}(b\verb`-`d)).  Certain traders, (such as Hedge Fund managers), may have very short trading horizons while others, (such as Pension Fund managers), have much longer horizons.  By looking at the value of the largest eigenvalue at different scales, we try to characterise the impact of these different trading horizons on the cross-correlation dynamics between large capitalisation stocks.  In Fig.~\ref{eValueLarge}(b\verb`-`d), we see that the main features found in the unfiltered data are preserved over longer time scales. However, certain features, such as the sizeable drop in the largest eigenvalue at the longest scale in late $2003$, are not seen at shorter scales, but the aggregate impact for the unfiltered data is a moderate drop.  Other features, such as the increase in $2006$, are not preserved across all scales.  The different features, found at various scales, suggest that the correlation matrix is made up of interactions between stocks, traded by investors with different time horizons.  This has implications for risk management, as the correlation matrix used for input in a portfolio optimisation should depend on the investor's time horizon.  We look more closely at this in Section~\ref{portOpt}.

\subsection{High-Frequency Correlation Analysis}
\label{HFcorrAnal}
Using the entire high-frequency data set for the Euro Stoxx $50$, described earlier, we calculated the average correlation and eigenvalue spectrum at a number of scales, Table~\ref{Table1}.  As expected, the largest eigenvalue increases as the average correlation between stocks increases, in keeping with the `one-factor' toy-model of correlations described previously, \cite{Conlon_2009}.  Analysis of Table~\ref{Table1} gives some intuition for the results shown, Fig.~\ref{eValueLarge}, with increases in the largest eigenvalue in time corresponding to an increase in the average correlation between stocks, (`single factor' model, \cite{Conlon_2009}).

\begin{table*}[htbp!]

	\centering
		\begin{tabular}{c c c | c c c c}
				& & \emph{Time}  &  \emph{Average} & \emph{$1^{st}$}  &  \emph{$2^{nd}$}  & \emph{$3^{rd}$} \\
				& \emph{Scale} & \emph{Horizon}  &  \emph{Correlation} & \emph{Eigenvalue}  &  \emph{Eigenvalue}  & \emph{Eigenvalue} \\
				\hline
				\hline
				& \emph{1} & \emph{1 min} & 0.35 & 20.24 & 6.66	& 1.085 \\
				& \emph{2} & \emph{5 min} & 0.39 & 20.66 & 5.53	& 1.26 \\
				& \emph{3} & \emph{30 min} & 0.45 & 23.33 & 4.229	& 1.64 \\
				& \emph{5} & \emph{120 min} & 0.47 & 24.7  & 4.18	& 1.74 \\
				& \emph{8} & \emph{480 min} & 0.52 & 27.72  & 3.54	& 2.42 \\
				& \emph{10} & \emph{1100 min} & 0.51 & 27.15  & 3.83	& 1.97 \\
				& \emph{12} & \emph{2000 min} & 0.41 & 18.42 & 2.78	& 2.39	\\			
				\hline									
		\end{tabular}
		\caption{Correlation and Eigenspectrum Analysis.  The $1^{st}$ eigenvalue is the largest eigenvalue of the correlation matrix of Euro Stoxx $50$ high frequency data from May $2008$ to April $2009$.}
			\label{Table1}
\end{table*}

As mentioned, (Section~\ref{Intro}), the Epps effect is the gradual build-up of correlation as the scale at which stock returns are measured, increases.   In previous work, \cite{Kertesz_2006,Kertesz_2007_a}, the Epps effect has been studied using either the bivariate correlation between pairs of stocks or the average of a number of such pairs.  However, as seen in Table~\ref{Table1}, this effect is also visible using a multivariate approach, with the largest eigenvalue increasing from $20.24$ at the smallest scale to $27.72$ at the $480$ min ($\approx1$ day) scale, with corresponding increase in average system correlation (from $0.35$ to $0.52$).    

In previous studies, Authors have calculated the correlation at different scales by recalculating the returns, resulting in reduced quantities of available data, and constraining the maximum possible scale.  However, using Wavelet Multiscaling to decompose the data into component scales, we can calculate the correlations and eigenvalues at much longer scales.  Of particular interest is the increase in the average correlation to a maximum at a scale of $480$ minutes and the subsequent decrease at longer scales of $2000$ min ($\approx 4$ days) to $0.41$.  This decrease in correlation corresponds to a decrease in the `Market Effect', with traders having longer time-horizons acting in a less synchronous fashion.  The decrease may be due to the additional discontinuities incorporated in the data at scales longer than one day as additional information is incorporated as stock markets open for the day, (visible in discontinuous jumps in the equity prices).  
				
\subsection{High-Frequency Correlation Dynamics}
\label{HFcorrDyn}
\begin{figure}[htbp!]
\begin{center}
\includegraphics[height=85mm,width=135mm]{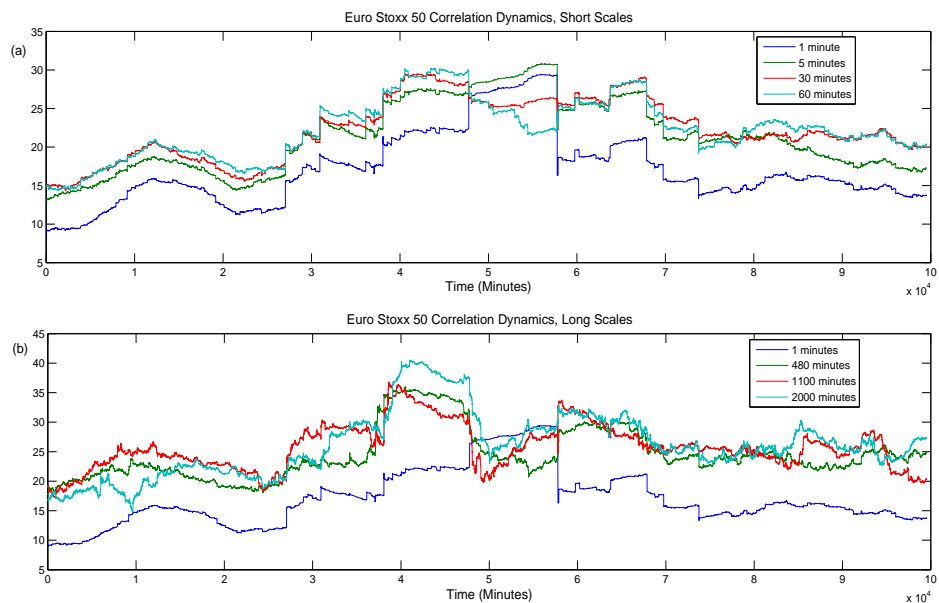}
\caption{Eigenvalues dynamics}
\label{eHFLarge}
\end{center}
\end{figure}

Building on the analysis of previous Sections, we now examine the dynamics of the eigenvalues over time using the high-frequency data described.  The methodology is the same as for daily data, described in Section~\ref{MedFreqDyn}, using a time window of $10,000$ minutes.  The results for small scales are shown, Fig.~\ref{eHFLarge}(a), with a general increase in the largest eigenvalue for longer scales, corresponding to the increase in correlation shown in Section~\ref{HFcorrAnal}.  However, for small scales ($1-5$ minutes), some distinct dynamics are found, with a sharp increase at around $50,000$ minutes and a decrease at longer scales ($30-60$ minutes).  These distinct dynamics emerge even more distinctly at longer scales, Fig.~\ref{eHFLarge}(b), with eigenvalues actually having opposite behaviour at certain points, (the largest eigenvalue increases at the $1$ minute scale but actually decreases at the longest scales, at approximately $50,000$ minutes).  

Of particular interest is the gradual increase in the largest eigenvalue moving to longer scales, in keeping with the Epps effect.  As found in the previous Section at longer scales, ($2,000$ minutes), using the complete data set, the largest eigenvalue is smaller than for shorter scales at certain points in time.  This variation in the largest eigenvalue corresponds to a change in the average correlation between stocks.  The distinct behaviour at longer time scales may be due to a number of factors, with longer scales incorporating additional information and smoothing out high frequency noise.

\begin{figure}[htbp!]
\begin{center}
\includegraphics[height=85mm,width=135mm]{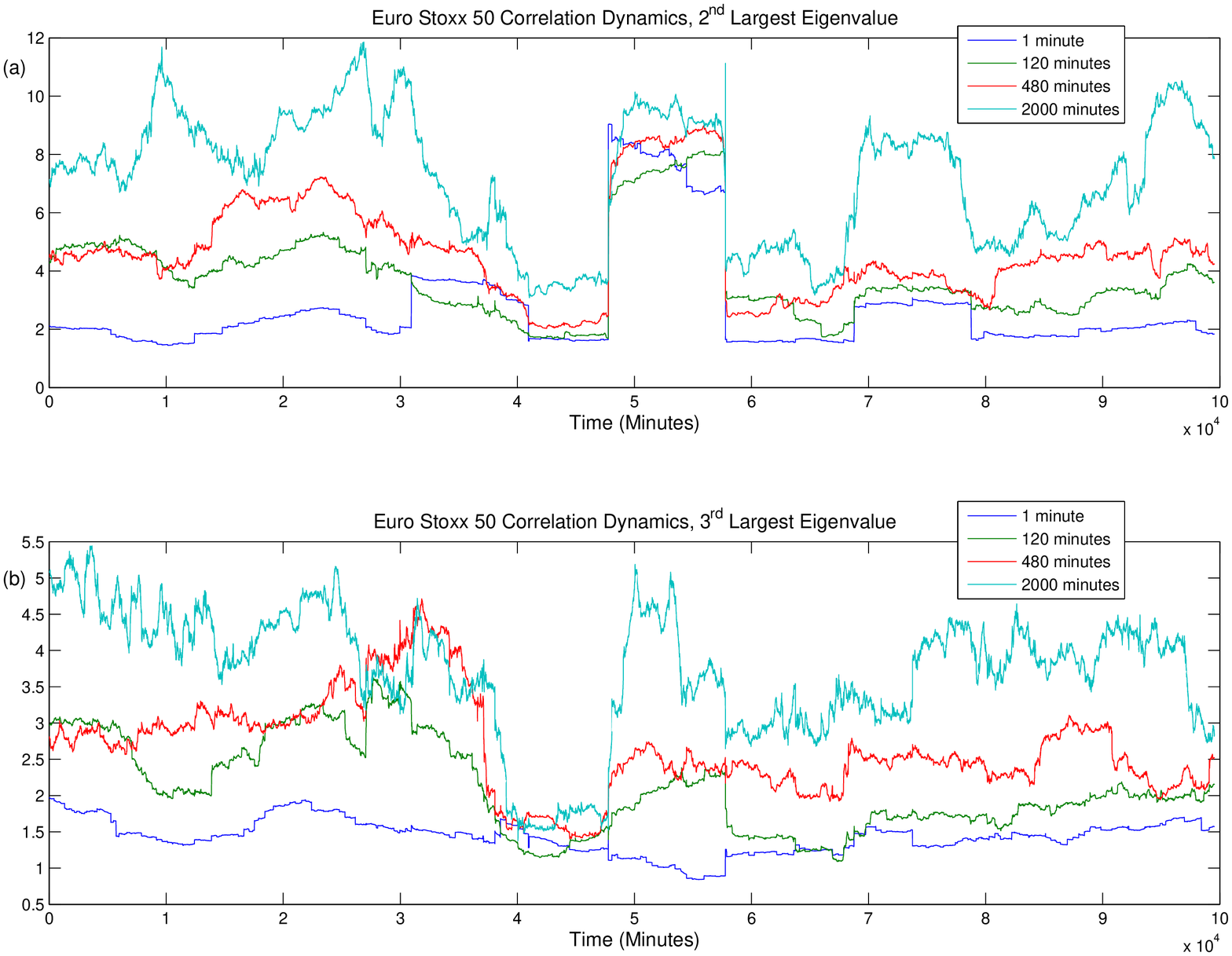}
\caption{Eigenvalues dynamics, $2^{nd}$ and $3^{rd}$ largest eigenvalues}
\label{eOthersLarge}
\end{center}
\end{figure}

The dynamics of the $2^{nd}$ and $3^{rd}$ largest eigenvalue are also shown in Fig.~\ref{eOthersLarge}.  Again, the relative size and dynamics of the eigenvalues are found to be scale-dependent, with a broad increase in the eigenvalues for longer scales.   This is in sharp contrast to the results found in Section~\ref{HFcorrAnal}, where the $2^{nd}$ largest eigenvalue was found to strictly decrease as the scale increased.  It appears that for small windows, the \emph{second largest eigenvalue is of greater relative importance} (at longer scales) with a corresponding redistribution of correlation structure across the eigenspectrum, (eigenvalue repulsion).  In the context of previous work done on Random Matrix Theory, \cite{Plerou_2000,Conlon_2007}, this may imply that different information is to be found in the $2^{nd}$ and subsequent eigenvalues at longer scales.  These eigenvalues are said to correspond to different sectoral groupings, with possible group changes as we move down the scales.  By removing the market mode, researchers \cite{Marsili_2007}, found the correlation structure to be invariant across scales.  However, this paper did not examine different time-windows or longer time-scales ($> 255$ Mins).  By combining the method of these authors with those proposed above, it may be possible to investigate the invariance of correlations across time as well as scale.

\subsection{Drawdown analysis}
As indicated, (Section~\ref{Intro} and \cite{Conlon_2009,Drozdz_2000}), drawdowns, (or periods of large negative returns), and drawups, (periods of large positive returns), tend to be accompanied by an increase in different eigenstates of the cross-correlation matrix. In this section, we attempt to characterise the market according to the relative size of the eigenvalues, as well as through the use of eigenvalue ratios. 

The returns, correlation matrix and eigenvalue spectrum time-series for overlapping windows of $10,000$ minutes were calculated and normalised using the mean and standard deviation over the entire series, (Eqn.~\ref{normalise}).  By representing normalised eigenvalues in terms of standard deviation units (SDU), we can partition the eigenvalues according to their magnitude.  The average return of the index is shown in Table~\ref{evalueReturns}, during periods when the largest eigenvalue is $\pm1$ SDU).  As the market trend was predominantly negative, average returns were negative both for periods when the largest eigenvalue was equal to its mimimum and maximum values.  However, distinct behaviour still emerged across all scales.

\begin{table*}[htbp!]
	\centering
		\begin{tabular}{c  c | c c c}
		\hline
				& \emph{Scale} & \emph{No. Std}  &  \emph{$>$ 1} &  \emph{$<$ -1} \\
		\hline
		\hline
				& \emph{Original} & &	-8.06\% & -3.39\% \\
				& \emph{5 Minutes} & & -7.95\% & -3.06\% \\
				& \emph{60 Minutes} & & -9.27\% & -3.64\% \\
				& \emph{360 Minutes} & & -8.41\% & -3.52\% \\
				& \emph{1100 Minutes} & & -10.79\% & 0.01\% \\
				& \emph{2000 Minutes} & & -12.06\% & -6.32\% \\ 
		\hline
		\hline
								
		\end{tabular}
		\caption{Drawdown/Drawup analysis. Average Index Returns when various eigenvalue partitions in SDU are $>1$ and $<-1$.}
			\label{evalueReturns}
\end{table*}

Looking first at the original or unfiltered data, we see that when the largest eigenvalue is $>1$ SDU, the return of the index was $-8.06\%$.  In contrast, when the largest eigenvalue is $< -1$ SDU, the return was $-3.06\%$.   Moving up the scales, similar behaviour was found with large eigenvalues being characterised by more pronounced downward moves.  The emergence of this characteristic at very small scales ($\approx 5$ minutes), implies that it is not a result of data granularity. 

\subsection{Portfolio Optimisation}
As mentioned previously, the time-scale dependence of correlation structure has implications for risk management.  In this Section we demonstrate, using high-frequency $1$ minute stock returns, that the optimal portfolio found from classic portfolio optimisation is time-scale dependent.  The correlation matrix between stocks of the Euro Stoxx $50$ was calculated for two time-frames, the first being a short time window of $10,000$ minutes, (corresponding to the first time window studied in Section~\ref{HFcorrDyn}), the second the complete data set of $109,540$ minutes, (corresponding to the data studied in Section~\ref{HFcorrAnal}).  The portfolio optimisation was unconstrained, allowing `short-selling' of stocks, which means that positive expected returns for the portfolio were possible, even though the majority of expected returns were negative.  The expected returns and risk for the equities were the same for both optimisations, meaning that all deviations in the risk/return profile are due to changes in the correlation structure.

\label{portOpt}

\begin{figure}[htbp!]
\begin{center}
\includegraphics[height=90mm,width=135mm]{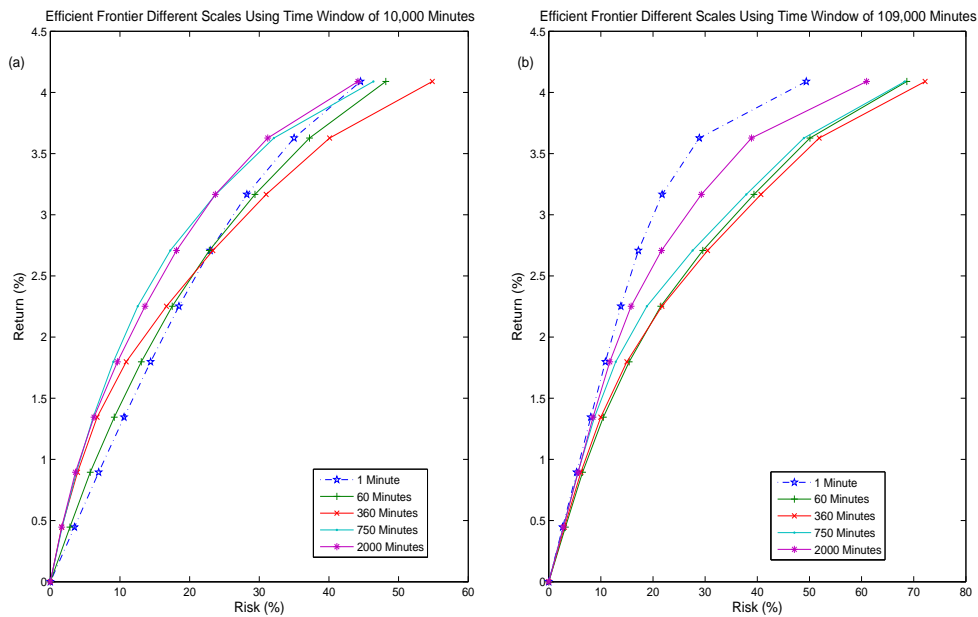}
\caption{Portfolio Optimisation}
\label{portOptGraph}
\end{center}
\end{figure}

The portfolio optimisation was performed in a standard fashion, \cite{Sharpe_1964,Elton_Gruber_2002,Bouchaud_book}, and the results, shown in Fig.~\ref{portOptGraph}, are highly dependent on the time-scale studied.  Interestingly, for the smaller time-window of $10,000$ minutes, the portfolios tended to be less risky at longer scales.  In contrast, using the larger time-window to calculate the correlations between stocks, the least risky portfolio was found using a $1$ minute scale.  Portfolios calculated at the $60-750$ minute scales were the riskiest, with longer-period scales found to be less risky.  This is in keeping with the findings of Section~\ref{HFcorrAnal}, where the average correlation peaked at a scale of $480$ minutes, meaning that further diversification benefits (lower correlation) can be found at the shortest and longest scales.

The optimal portfolio is clearly dependent on the time-scale used to calculate the correlation matrix, which in turn, is dependent on the time window used.  Obviously, this means that care must be taken by investors in the choice of both time-window and granularity used in the calculation of correlation matrices.

\section{Conclusions}
\label{conclusions}
Using a multivariate technique first applied to complex EEG seizure data, the correlation structure between medium and high-frequency financial time-series was studied.  Application of the Maximum Overlap Discrete Wavelet Transform extended the study of changes in the cross-correlation structure across both scale and time. 

\begin{enumerate}
\item
Using the MODWT and a sliding window, the dynamics of the largest eigenvalue of the correlation matrix were examined and shown to be time dependent at all scales, (using both medium and high-frequency equity returns).  Similar dynamics were visible across all scales, but with particular features markedly apparent at certain scales.  This suggests that the correlation matrix between equities consists of interactions caused by traders with different time horizons.

\item
Study of the $2^{nd}$ and $3^{rd}$ largest eigenvalues over both time and scale revealed a large increase in sectoral correlations for longer scales, using high-frequency data.  

\item
High frequency data were used to calculate the average correlation and largest eigenvalue for various scales.  The \emph{Epps} effect was found using this multivariate technique and analysis of longer scales revealed a drop in the average system correlation at scales beyond one day.

\item
A partition of the time-normalised eigenvalues demonstrated quantitatively the relationship between the size of the largest eigenvalue and the return of the index.  This feature emerged at very small scales ($< 5$ minutes), implying that this relationship is not due to the granularity of the data studied. 

\item
Finally, we looked at the problem of portfolio optimisation at various scales.  Results show that the optimal portfolio depends not only on scale but on the time-window used in the calculation of the correlation matrix.
\end{enumerate}

The correlation technique used in this paper to measure the interaction between agents suffers from the drawback of being linear and hence neglects any higher-order relationships.  Future work includes the application of non-linear information-theory based dependence measures, which will allow the detection of complex changes in synchronisation behaviour around extreme financial events.  These non-linear techniques may uncover further emergent features to those highlighted above and may result in signals that warn of likely future market turmoil.   The interaction of various agents with competing strategies and operating over different time scales, requires further investigation in order to improve understanding of complex phenomema such as financial crashes.  Additionally, while this financial work has benefited from parallels drawn with EEG system analysis, insights gained here may also raise questions with regards to the behaviour of other complex systems.

\end{document}